\documentstyle[12pt]{article}

\def\be{\begin{equation}}
\def\ee{\end{equation}}
\def\br{\begin{eqnarray}}
\def\er{\end{eqnarray}}
\def\nonu{\nonumber}
\def\br{\begin{eqnarray}}
\def\er{\end{eqnarray}}

\def\ra{\rightarrow}
\def\no{\nonumber}
\def\bdi{\begin{displaymath}}
\def\edi{\end{displaymath}}

\def\RR{{\rm I\kern-.1567em R}}                              
 \def\CC{{\rm C\kern-4.7pt                                    
 \vrule height 7.7pt width 0.4pt depth -0.5pt \phantom {.}}} 
 \def\ZZ{{\sf Z\kern-4.5pt Z}}                                

\begin{document}

\begin{titlepage}
\vspace*{-2 cm}
\noindent

\vskip 3cm
\begin{center}
{\Large\bf The symmetries of the Dirac--Pauli equation in two and three 
dimensions}
\vglue 1  true cm
C. Adam\footnote{email: adam@radon.mat.univie.ac.at} and  
J. S\'anchez-Guill\'en\footnote{email: joaquin@fpaxp1.usc.es}
\vspace{1 cm}

\vspace{1 cm}
{\footnotesize Departamento de F\'\i sica de Part\'\i culas,\\
Facultad de F\'\i sica\\
Universidad de Santiago\\
and \\
Instituto Galego de Fisica de Altas Enerxias (IGFAE) \\
E-15706 Santiago de Compostela, Spain}

\medskip
\end{center}

\normalsize
\vskip 0.2cm

\begin{abstract}
We calculate all symmetries of the Dirac-Pauli equation in two-dimensional
and three-dimensional Euclidean space. Further, we use our results for
an investigation of the issue of zero mode degeneracy. We construct explicitly 
a class of multiple zero modes with their gauge potentials.

\end{abstract}

\vskip 2cm

\end{titlepage}

\section{Introduction}

The Abelian Dirac equation in Euclidean space in two and three dimensions is
\be \label{Dir-eq}
{\cal D} \Psi \equiv \sigma^k (-i\partial_k -G^k ({\bf r}))\Psi ({\bf r}) =0,
\ee
where, in three dimensions, 
${\bf r}= (x ,y,z)^{\rm t}$, $k =x,y,z$, $\sigma^k$ are the Pauli matrices,
and $G^k \equiv (G^x ,G^y ,G^z)$ is the gauge
potential (we use superscripts for the components because subscripts
will be exclusively reserved to describe partial derivatives). In two
dimensions all $z$ components are absent. Further,
$\Psi =(\Phi ,\chi )^{\rm t}$ is a two-component spinor.

The Pauli equation is obtained from the the Dirac equation by simply
squaring the Dirac operator ${\cal D}$ acting on $\Psi$,
\be
\left( (-i\partial_k -G^k )^2 -\sigma^l H^l \right) \Psi =0,
\ee
where $\vec H$ is the magnetic field, $\vec H =\nabla \times \vec G$. 
Solutions of the Dirac equation (\ref{Dir-eq}) are, at the same time,
solutions of the Pauli equation, therefore we shall treat them on an
equal footing.

Solutions to the  Pauli or Dirac equation are relevant in several
instances. They describe the behaviour of non-relativistic electrons
in the presence of magnetic fields, and their existence, among other
issues, influences
the stability of non-relativistic matter \cite{FLL}, \cite{LY1}. 
On the other hand, they
influence the behaviour of the Fermion determinant ${\rm det}\, {\cal D}$
for relativistic electrons and are, therefore, relevant for the
strong field behaviour of relativistic electrons
\cite{Fry1}--\cite{Fry3}, and for a proper
path-intergral quantization of QED in two \cite{Jay}--\cite{Ada1}
and three dimensions.

For the Dirac equation in two dimensions the most important information 
on solutions is provided by the Aharanov--Casher theorem
\cite{AC}, which states 
that there are $n$ square-integrable solutions when the magnetic flux  
divided by $2\pi$ is between $n$ and $n+1$, i.e., 
$n< ({\rm flux} /2\pi )\le n+1$. 
For the Dirac equation in
three dimensions the available information is much scarcer. The first 
examples of solutions have been given in 1986, see \cite{LY1}.
Further examples have bee  provided in \cite{AMN1}, \cite{El1}, \cite{AMN2},
and the existence of multiple solutions (zero mode degeneracy) has been
first demonstrated in \cite{AMN3}, \cite{AMN4}, and \cite{ES}. 
Further results on zero-mode supporting gauge potentials may be found in
\cite{BE1}, \cite{BE2}, and in \cite{El2}. A general classification of
zero-mode supporting gauge fields is still missing.

In this paper we study the symmetries of the Dirac equation (\ref{Dir-eq})
both in two and three dimensions, in order to gain some further insight
into the solution space of these equations. In Section 2 we calculate
the symmetries of the Dirac equation in two dimensions and briefly apply 
our results to the issue of multiple zero modes.
In Section 3 we calculate the symmetries of
the Dirac equation in three dimensions. We use the 
method of prolongations for all our symmetry calculations, which is described
in detail in the book \cite{Olv}. In Section 4,  we use our results on 
symmetries of the three-dimensional Dirac equation for
a discussion of the issue of zero mode degeneracy (i.e., multiple
solutions) in three dimensions. In Subsection 4.1, we discuss how our
results on symmetries can be used to construct multiple zero modes in
principle. In Subsection 4.2, we apply the results of Subsection 4.1
for the explicit construction of a class of multiple zero modes 
together with their corresponding gauge potentials.

\section{Symmetries of the Dirac equation in two dimensions} 

For the Dirac equation in two dimensions it is known from the Aharanov--Casher
theorem that
all zero modes (i.e., square-integrable solutions)
are either left-handed (i.e., the
lower component of $\Psi$ is zero) or right-handed (the upper component of
$\Psi$ is zero). Further, a solution of the first type (left-handed) may be 
mapped into a solution of the second type by the simple replacement
$G^k \ra -G^k$, therefore we may restrict, e.g.,  to the left-handed case
\be \label{dirac2}
(-i\partial_x -G^x +\partial_y -iG^y )\Phi =0.
\ee
By introducing the modulus and phase of $\Phi$ via $\ln \Phi = \rho +i\lambda
$ we may rewrite Equation (\ref{dirac2}) in terms of two real first order
equations as
\br \label{di-1}
\Delta_1 \; \equiv \; \, \rho_y + \lambda_x -G^x &=&0 \\
\Delta_2 \; \equiv \; \, \rho_x -\lambda_y +G^y &=&0 \label{di-2}
\er
where from now on subscripts denote partial derivatives, i.e., $\rho_x
\equiv \partial_x \rho$, etc.  

The vector field generating generic transformations on the independent and
dependent variables is 
\be
v=X \partial_x +Y\partial_y +R\partial_\rho +L\partial_\lambda +
F^x \partial_{G^x} +F^y \partial_{G^y}
\ee
where $X,Y,R,L,F^x ,F^y$ may depend on all independent and dependent 
variables. The equations (\ref{di-1}, \ref{di-2}) contain first derivatives 
of the
variables $\lambda $ and $\rho$, therefore we need the first prolongation
of $v$ w.r.t. these variables,
\be
{\rm pr}\, v    = v + R^x \partial_{\rho_x} +R^y \partial_{\rho_y} 
 + L^x \partial_{\lambda_x} +L^y \partial_{\lambda_y},
\ee
where, e.g.
\be
R^x  = D_x R -\rho_x D_x X - \rho_y D_x Y .
\ee
Here $D_x$ is a total derivative, and we give the explicit (rather 
lengthy) expressions in  
Appendix A.

Now, a symmetry of a PDE $F(\rho ,\rho_j ,\ldots )=0$ (of $n$-th order, say)
is a solution of the equation ${\rm pr}^{(n)} v(F)=0$ which holds on-shell,
i.e., when the original PDE is used together with its prolongations
(PDEs that follow from $F=0$ by applying total derivatives). As said,
for the details of the used formalism we refer to the book (\cite{Olv}).

Concretely, the determination of the symmetry transformation proceeds 
as follows. We 
require that ${\rm pr}\, v(\Delta_1) =0 ={\rm pr}\, v(\Delta_2)$ whenever the
equations (\ref{di-1}) and (\ref{di-2}) hold. Explicitly this means that
\br
R^y +L^x -F^x &=& 0 \label{co-1} \\
R^x -L^y +F^y &=& 0 \label{co-2}
\er 
whenever the equations  (\ref{di-1}) and (\ref{di-2}) hold. Equations
(\ref{co-1}) and (\ref{co-2}) contain a number of algebraically independent
functions (like $\rho_x$, $\rho_x G^y_x$, etc.) multiplying the coefficients
(like $R_x$, $R_\rho$, etc.) which we want to determine. We may use the
equations  (\ref{di-1}) and (\ref{di-2}) to eliminate, e.g., $\lambda_x$
and $\lambda_y$ from the equations (\ref{co-1}) and (\ref{co-2}). Then we 
demand that the coefficient of each algebraically independent function
(like $\rho_x$, $\rho_x G^y_x$, etc.) vanishes separately. This leads to
more than 30 conditions for each of the two equations 
(\ref{co-1}) and (\ref{co-2}).
Fortunately, most of these conditions are quite trivial, like, e.g., 
$X_\lambda =0=X_{A^x}$, etc. The final result is that Eq. (\ref{co-1})
leads to the conditions 
\be
X=X(x,y) \, ,\quad Y=Y(x,y)\, ,
\ee
\be
R=R(\rho ,\lambda ,x,y) \, ,\quad L=L(\rho ,\lambda ,x,y) \, ,
\ee
\br
R_\rho -L_\lambda  &=&  Y_y - X_x  \\
L_\rho +R_\lambda &=&  X_y + Y_x 
\er
and
\be
F^x =L_x +R_y +( L_\lambda -X_x )G^x + (R_\lambda -Y_x )G^y 
\ee
whereas Eq. (\ref{co-2}) leads to the conditions
\be
X=X(x,y) \, ,\quad Y=Y(x,y)\, ,
\ee
\be
R=R(\rho ,\lambda ,x,y) \, ,\quad L=L(\rho ,\lambda ,x,y) \, ,
\ee
\br
R_\rho -L_\lambda  &=&  -Y_y + X_x  \\
L_\rho +R_\lambda &=&  -X_y - Y_x 
\er
and
\be
F^y = L_y -R_x -(R_\lambda +X_y) G^x + (L_\lambda -Y_y )G^y .
\ee
For later convenience we 
introduce the complex notation
\be
z\equiv x+iy \, , \quad
Z\equiv X+iY \, , \quad
\sigma \equiv \rho +i\lambda \, ,\quad 
 \Sigma \equiv R +iL
\ee
and
\be
G \equiv \frac{1}{2} (G^x +iG^y) \, ,\quad F \equiv \frac{1}{2}
(F^x +iF^y)
\ee
which implies 
\be
\partial_\sigma \equiv \frac{1}{2}(\partial_\rho -i\partial_\lambda )
\ee
etc.
Compatibility between the two sets of equations requires
\be
X_x =Y_y \, \quad X_y =-Y_x
\ee
which are just the Cauchy--Riemann equations, and 
\be
L_\rho =L_\lambda \, ,\quad R_\lambda =-L_\rho \, ,
\ee
which are the Cauchy--Riemann equations w.r.t. the target space variable
$\sigma = \rho +i\lambda$.
Therefore, $X$ and $Y$ are the real and imaginary part of a holomorphic 
function,
\be
X+iY = Z(z)\, ,\quad z=x+iy
\ee
whereas $L$ and $R$ are the real and imaginary part of a holomorphic function
in the variable $\sigma$, 
\be
R+iL = \Sigma (\sigma ,z ,\bar z),
\ee
where $\Sigma$ may still depend on $z,\bar z$.
Finally, for $F$ we find
\be \label{F-2d}
F\equiv \frac{1}{2} (F^x +iF^y) =
-i \Sigma_{\bar z} -G \bar Z_{\bar z} +G\Sigma_\sigma .
\ee
The most general symmetry generator
\be v=Z\partial_z +\bar Z\partial_{\bar z} +\Sigma\partial_\sigma
+\bar \Sigma \partial_{\bar\sigma} +F\partial_G +\bar F \partial_{\bar G}
\ee
may be expressed as the semi-direct sum of a symmetry generator 
$v_Z$ w.r.t. the
holomorphic function $Z(z)$ and a generator $v_\Sigma$ w.r.t. 
the function $\Sigma (\sigma ,z ,\bar z)$, where
\be
v_Z = Z\partial_z +\bar Z \partial_{\bar z} 
-G \bar Z_{\bar z}  \partial_{ G} - \bar G Z_z \partial_{\bar G}
\ee
and
\be
v_\Sigma = \Sigma \partial_\sigma +\bar\Sigma \partial_{\bar \sigma}
+ (- i \Sigma_{\bar z} +  G  \Sigma_{ \sigma} )\partial_{ G}
+ (i\bar \Sigma_z + \bar G \bar \Sigma_{\bar \sigma} )\partial_{\bar G} 
 \, .
\ee
Further, they obey the infinite-dimensional Lie algebra
\be
[v_{Z_1},v_{Z_2}] = v_{Z_3} \, , \quad Z_3 =Z_1 Z_2' -Z_2 Z_1' \, ,
\ee
\be
[v_{\Sigma_1},v_{\Sigma_2}] =
v_{\Sigma_3} \, , \quad \Sigma_3 =\Sigma_1 \Sigma_{2,\sigma} -\Sigma_2 
\Sigma_{1,\sigma} \, ,
\ee
which are just two copies of the Virasoro algebra in coordinate and
target space, and
\be
[v_Z, v_\Sigma ] = v_{\tilde \Sigma} \, ,\quad \tilde \Sigma = 
\Sigma_z Z + \Sigma_{\bar z} \bar Z \, .
\ee
It is quite interesting that the symmetry turns out to be so large, the
semi-direct product of two conformal groups in two dimensions.
    
[Remark: the symmetry transformations we found cover the whole solution space 
in the sense that any solution $\Phi$ of the Dirac equation (\ref{dirac2})
for any gauge potential $G$ can be found by applying a symmetry transformation
to the trivial solution $\Phi =1$, $G=0$. In fact, a target space 
transformation $\Sigma$ of the type $\Sigma = f(z ,\bar z)$ is sufficient.
For this sub-class of transformations the above Lie algebra is Abelian,
therefore the exponentiation for finite $\Sigma$ is trivial.]

As an application let us briefly discuss the issue of degeneracy of zero
modes (i.e., multiple solutions of the Dirac equation for one gauge field
$G$). The condition that the gauge field does not change is, of course,
that the $F$ in Eq. (\ref{F-2d}) is zero, which leads to the conditions
$\Sigma_\sigma = \Sigma_{\bar z} =0$, that is $\Sigma =g(z)$ is a
function in the variable $z$ only. As said, the exponentiation
of such $\Sigma$ is trivial, therefore a
\be
\Phi ' =e^{g(z)} \Phi
\ee
is a local zero mode for the same gauge potential as $\Phi$. Single-valuedness
of $\Phi '$ on the whole Euclidean plane restricts the allowed $g$
($\exp g(z)$ should be single-valued) and the condition of square-integrability
further restricts the allowed $g(z)$. The fact that further zero modes
of the same Dirac operator may be produced by multiplying the given one
with analytic functions of $z$ is, of course, well-known.

\section{Symmetries of the Dirac equation in three dimensions}

Introducing the real and imaginary part of the spinor components via
$\Phi = \alpha +i\beta$, $\chi = \gamma + i\delta$, the Dirac equation
(\ref{Dir-eq}) in three dimensions
is equivalent to the four real equations
\br \label{Deq-3d}
I \quad \equiv \quad
\beta_z +\delta_x -\gamma_y -G^z \alpha -G^x \gamma -G^y \delta &=& 0 \no \\
II \quad \equiv \quad
\alpha_z +\gamma_x +\delta_y +G^z \beta +G^x \delta -G^y \gamma &=& 0 \no \\
III \quad \equiv \quad
\beta_x +\alpha_y -\delta_z -G^x \alpha +G^y \beta +G^z \gamma &=& 0 \no \\
IV \quad \equiv \quad
\alpha_x - \beta_y -\gamma_z +G^x \beta +G^y \alpha -G^z \delta &=& 0 .
\er
If viewed as a system of algebraic equations for the three real unknowns
$(G^x ,G^y ,G^z)$, these four equations must be algebraically dependent.
The dependence is given by the condition
\be \label{constr}
V \, \equiv \,
\nabla \cdot \vec \Sigma = 2(\alpha \gamma +\beta \delta )_x
+2 (\alpha \delta - \beta \gamma )_y + (\alpha^2 +\beta^2 -\gamma^2 
-\delta^2 )_z =0
\ee
where
\be
\vec \Sigma \equiv \Psi^\dagger \vec\sigma \Psi =
\left( \begin{array}{c} 2(\alpha \gamma + \beta \delta ) \\
2(\alpha \delta - \beta \gamma ) \\ \alpha^2 +\beta^2 -\gamma^2 - \delta^2
\end{array} \right)
\ee
is the spin density of the spinor $\Psi$.  

Again, we want to study the symmetries of the above equations (\ref{Deq-3d})
using
the method of prolongations. The symmetry generating vector field is
\be
v=X \partial_x +Y\partial_y +Z \partial_z + A\partial_\alpha +
B\partial_\beta + C\partial_\gamma + D\partial_\delta +
F^x \partial_{G^x} +F^y \partial_{G^y} + F^z \partial_{G^z}
\ee
and its prolongation to the needed order is
\br \label{prol-3d}
{\rm pr} \, v &=& v + A^x \partial_{\alpha_x} +A^y \partial_{\alpha_y} 
 + A^z \partial_{\alpha_z} +
B^x \partial_{\beta_x} +B^y \partial_{\beta_y} 
 + B^z \partial_{\beta_z} \nonumber \\
&& + C^x \partial_{\gamma_x} +C^y \partial_{\gamma_y} 
 + C^z \partial_{\gamma_z} +
D^x \partial_{\delta_x} +D^y \partial_{\delta_y} 
 + D^z \partial_{\delta_z}
\er
where, e.g., $A^x$ is given by
\be \label{A^x}
A^x = A_x +A_\alpha \alpha_x +A_\beta \beta_x + A_\gamma \gamma_x + A_\delta
\delta_x - \alpha_x X_x - \alpha_y Y_x - \alpha_z Z_x
\ee
Here, we have already used some first results of the calculation below, namely
that (analogously to the case in two dimensions) 
no coefficient depends on the gauge potential, and that the coefficients
$X,Y,Z$ of the translations in base space only depend on the base space
coordinates $x,y,z$ (otherwise the resulting expression (\ref{A^x})
would be more complicated, analogously to the $R^x$ of the 
two-dimensional case, which is given in Appendix A). 

Acting with the prolonged vector field on the four equations (\ref{Deq-3d}) 
(${\rm  pr} \, v (I) =0$, etc.) leads to the
following set of four equations
\br \label{pr-Deq-3d}
 G^x C +G^y D + G^z A +F^x \gamma + F^y \delta 
+F^z \alpha - D^x +C^y - B^z &=& 0 \no \\
 G^x D - G^y C + G^z B +F^x \delta - F^y \gamma 
+ F^z \beta + C^x + D^y + A^z &=& 0 \no \\
 -G^x A + G^y B + G^z C - F^x \alpha + F^y \beta 
+ F^z \gamma + B^x + A^y - D^z &=& 0 \no \\
 G^x B + G^y A - G^z D + F^x \beta + F^y \alpha 
- F^z \delta + A^x - B^y - C^z &=& 0 \no \\ &&
\er 
Next we have to insert the explicit expressions for $A^x$ etc. We obtain 
algebraic expressions in terms of the target space variables and their partial
derivatives, where we may eliminate four derivative terms with the help
of the Dirac equation (\ref{Deq-3d}).
(Explicitly, we eliminated $\delta_x ,\delta_y ,\delta_z
$ and $\gamma_z$.) For the resulting expressions one has to require that
the coefficient multiplying each partial derivative of the target space
variables (like $\alpha_x$ or 
$G^x_x$) or product of partial derivatives vanishes separately. Doing this one
realizes quickly that nothing may depend on the gauge potential and that
the coefficients $X,Y,Z$ may only depend on $x,y,z$. 

The remaining coefficients which are multiplied by at least one partial
derivative of a target space variable serve to determine the coefficients
$X,Y,Z$ and $A,B,C,D$. They are of two types. The first type consists
of pairs of equations like 
\bdi
A_\delta + D_\alpha = Y_z +Z_y \, ,\quad 
 A_\delta + D_\alpha = -Y_z - Z_y \quad \Rightarrow 
A_\delta + D_\alpha = 0 \, ,\quad Y_z +Z_y =0
\edi
which lead to equations for the $X,Y,Z$ and $A,B,C,D$ independently. The 
equations for $X,Y,Z$ are
\bdi
X_x =Y_y =Z_z
\edi
\be
X_y +Y_x =0 \, ,\quad X_z +Z_x =0 \, ,\quad Y_z +Z_y =0
\ee
and have the conformal transformations in three-dimensional space as general
solutions,
\br \label{conf-3d}
X &=& \frac{\theta^1}{2}(x^2 -y^2 -z^2 ) +\theta^2 xy +\theta^3 xz 
-l^3 y +l^2 z +\sigma x +x^0 \no \\
Y &=& \frac{\theta^2}{2}(y^2 -x^2 -z^2)+\theta^1 xy +\theta^3 yz 
+l^3 x -l^1 z + \sigma y +y^0 \no \\
Z &=& \frac{\theta^3}{2} (z^2 -x^2 -y^2) +\theta^1 xz +\theta^2 yz 
-l^2 x+l^1 y +\sigma z +z^0 
\er  
Here $\theta^i , l^i ,\sigma$ and $\vec r^0$ are constant parameters 
which parametrize the infinitesimal conformal transformations.

The equations for $A,B,C,D$ are
\bdi
A_\alpha =B_\beta = C_\gamma =D_\delta
\edi
\bdi
A_\beta +B_\alpha =0 \, ,\quad A_\gamma +C_\alpha =0 \, ,\quad 
A_\delta + D_\alpha =0 
\edi
\be
B_\gamma +C_\beta =0 \, ,\quad B_\delta +D_\beta =0 \, ,\quad
C_\delta +D_\gamma =0
\ee
and have the conformal transformations in the four-dimensional 
target space as solutions,
\br \label{conf-4d}
A &=& \frac{\zeta^1}{2}(\alpha^2 -\beta^2 -\gamma^2 -\delta^2) +\zeta^2
\alpha \beta +\zeta^3 \alpha \gamma +\zeta^4 \alpha \delta + \no \\
&& \lambda \alpha -v^3 \beta + v^2 \gamma -v^1 \delta +\alpha^0 \no \\
B &=& \frac{\zeta^2}{2}(\beta^2 -\alpha^2 -\gamma^2 -\delta^2) +\zeta^1
\alpha \beta +\zeta^3 \beta\gamma +\zeta^4 \beta\delta + \no \\
&& \lambda\beta +v^3 \alpha -u^1 \gamma -u^2 \delta +\beta^0 \no \\
C &=& \frac{\zeta^3}{2}(\gamma^2 -\alpha^2 -\beta^2 -\delta^2) +\zeta^1 
\alpha\gamma +\zeta^2 \beta\gamma +\zeta^4 \gamma\delta + \no \\
&& \lambda\gamma -v^2 \alpha + u^1 \beta -u^3 \delta +\gamma^0 \no \\
D &=& \frac{\zeta^4}{2} (\delta^2 -\alpha^2 =\beta^2 -\gamma^2) +\zeta^1
\alpha\delta +\zeta^2 \beta \delta +\zeta^3 \gamma\delta + \no \\
&& \lambda \delta +v^1 \alpha +u^2 \beta + u^3 \gamma + \delta^0
\er 
Here the parameters $\zeta^j , u^i ,v^i , \lambda$ and $\alpha^0 ,\beta^0 
,\gamma^0 ,\delta^0$ may still depend on the base space coordinates $(x,y,z)$.
 
The second type of equations consists of equations like $C_\delta -A_\beta
= X_y$ and establishes relations between the $A,B,C,D$ and the $X,Y,Z$,
thereby further restricting the possible $A,B,C,D$. The result is that the
proper conformal transformation on target space must be absent, $\zeta^j
\equiv 0$, and that the six target space rotation parameters  are no longer
independent,
\br
u^1 -v^1 &=& l^1 +\theta^2 z - \theta^3 y \equiv L^1 \no \\
u^2 -v^2 &=& l^2 +\theta^3 x - \theta^1 z \equiv L^2 \no \\
u^3 -v^3 &=& l^3 +\theta^1 y - \theta^2 x \equiv L^3 
\er
leading to
\br \label{A-res1}
A &=& \lambda \alpha -v^3 \beta + v^2 \gamma -v^1 \delta +\alpha^0 \no \\
B &=& \lambda\beta +v^3 \alpha -(v^1 + L^1) \gamma -(v^2 + L^2) 
\delta +\beta^0 \no \\
C &=& \lambda\gamma -v^2 \alpha + (v^1 + L^1) \beta -
(v^3 + L^3) \delta +\gamma^0 \no \\
D &=& \lambda \delta +v^1 \alpha + (v^2 + L^2) \beta + 
(v^3 + L^3) \gamma + \delta^0
\er 
where $\lambda ,v^i$ and $\alpha^0 ,\ldots$ may still depend on $(x,y,z)$.
By shifting $v^i \to v^i -(L^i /2)$ the $L^i$ are distributed more
symmetrically,  
\br \label{A-res2}
A &=& \lambda \alpha -(v^3 -\frac{L^3}{2}) \beta + (v^2 - \frac{L^2}{2}) 
\gamma -(v^1 -\frac{L^1}{2}) \delta +\alpha^0 \no \\
B &=& \lambda\beta +(v^3 - \frac{L^3}{2}) \alpha -(v^1 + \frac{L^1}{2}) 
\gamma -(v^2 + \frac{L^2}{2}) 
\delta +\beta^0 \no \\
C &=& \lambda\gamma -(v^2 - \frac{L^2}{2}) \alpha + (v^1 + \frac{L^1}{2}) 
\beta -
(v^3 + \frac{L^3}{2}) \delta +\gamma^0 \no \\
D &=& \lambda \delta +(v^1 - \frac{L^1}{2}) \alpha + (v^2 + \frac{L^2}{2}) 
\beta + 
(v^3 + \frac{L^3}{2}) \gamma + \delta^0
\er 
and we find the half-integer valued representation of the base space
rotations (with parameters $l^i$)
acting on the (spinor) target space variables. It is interesting 
to note that the half-integer values appear also for the parameters of the
proper conformal transformations on base space (remember $L^i \equiv
l^i +\epsilon^{ijk}\theta^j r^k$).

[Remark: for a Dirac equation without the additional condition $\nabla
\cdot \vec \Sigma =0$, see Eq. (\ref{constr}), the above expressions would 
be the final result. This is the case, for instance, for the Dirac
equation for a Weyl (i.e., two-component) spinor in Euclidean space in
four dimensions, for which expressions analogous to (\ref{A-res2}) 
would constitute the final result (with the conformal base space
transformations in 4 instead of 3 dimensions, of course).]

We have used all information from the coefficients of equations 
(\ref{pr-Deq-3d}) which
contain at least one partial derivative of a target space variable. It 
remains to evaluate the parts without a partial derivative (i.e., the
coefficients of the identity). These contain the coefficients $F^x ,F^y
,F^z$ linearly and are therefore just the determining linear equations for
these coefficients. However, they give 4 equations for 3 unkonwns, therefore
they must be linearly dependent, which leads to one further constraint
equation. An easier way to calculate this constraint is to calculate the
action of the prolonged vector field (\ref{prol-3d}) on the original 
constraint (\ref{constr}),
\be
{\rm pr} \, v (V )=0
\ee
or explicitly
\bdi
A^x \gamma +\alpha_x C +A \gamma_x +\alpha C^x + B^x \delta +\beta_x D +
B \delta_x + \beta D^x +
\edi
\bdi
A^y \delta + \alpha_y D +A \delta_y +\alpha D^y -B^y
\gamma -\beta_y C -B \gamma_y -\beta C^y +
\edi
\be \label{pr-constr}
A^z \alpha +\alpha_z A +B^z \beta 
+\beta_z B - C^z \gamma -\gamma_z C - D^z \delta -\delta_z D =0    
\ee
Inserting the explicit expressions for $A^x$, etc. leads to an expression
containing both coefficients multiplying partial derivatives of the target
space variables (like $\alpha_x$) and a coefficient of the identity 1. 
After eliminating four partial derivatives (e.g. $\delta_x ,\delta_y ,
\delta_z$ and $\gamma_z$) with the help of the e.o.m., one realizes that
the coefficients of the remaining partial derivatives vanish identically.
It remains to evaluate the coefficient of the identity. This leads in fact
to four conditions, because the use of the e.o.m. has reintroduced the
gauge potential functions $G^x ,G^y ,G^z$ into the above expression. As none
of the coefficients ($A$, etc.) may depend on the gauge potential, the total
coefficient multiplying each gauge potential function has to vanish 
separately, as well as the remainder. The conditions that the coefficients
mulitplying $G^x ,G^y ,G^z$ vanish leads to
\be
\alpha^0 = \beta^0 = \gamma^0 = \delta^0 =0 \, ,\quad
v^1 = v^2 =0
\ee
and the condition that the remainder vanishes leads to 
\be
(\nabla \lambda + \vec \theta ) \cdot \vec\Sigma =0
\ee
With the shift
\be
\lambda \to \lambda -\Theta \, ,\quad \Theta \equiv \vec \theta \cdot \vec r
\equiv \theta^1 x +\theta^2 y +\theta^3 z
\ee
and the definition
\be
\phi \equiv - v^3
\ee
we therefore arrive at
\br \label{A-fin}
A &=& \lambda  \alpha + \phi \beta - \Theta \alpha + \frac{L^3}{2} \beta 
 - \frac{L^2}{2}
\gamma + \frac{L^1}{2} \delta  \no \\
B &=& \lambda\beta  - \phi \alpha - \Theta \beta - \frac{L^3}{2} \alpha -
 \frac{L^1}{2}  \gamma - \frac{L^2}{2} \delta  \no \\
C &=& \lambda\gamma + \phi \delta - \Theta \gamma + \frac{L^2}{2} \alpha + 
 \frac{L^1}{2} \beta -
 \frac{L^3}{2} \delta  \no \\
D &=& \lambda \delta - \phi \gamma - \Theta \delta  - \frac{L^1}{2} \alpha + 
 \frac{L^2}{2} \beta + 
 \frac{L^3}{2} \gamma 
\er 
with
\be \label{cond}
(\nabla \lambda )\cdot \vec \Sigma =0.
\ee
Here $\lambda$ and $\phi$ are scale and gauge transformations on (spinor) 
target space, whereas the remainder is the representation on spinor space of
the infinitesimal base space symmetries.

However, there is a problem with the constraint (\ref{cond}). The constrained 
function $\lambda$ is still algebraically independent of $\vec \Sigma$
(i.e., of the $\alpha ,\beta$, etc.), therefore the constraint does not
contradict our basic assumptions. The problem is that the Lie algebra
of all infinitesimal transformations does not close on the constraint 
(\ref{cond}).
E.g., the commutator of a target space scale transformation and a base space
rotation about the z-axis leads to another target space scale transformation
which no longer obeys the constraint (\ref{cond})
\be
[v_\lambda ,v_{l^3}] =v_{\tilde \lambda} \, ,\quad (\nabla \tilde \lambda )
\cdot \vec \Sigma \ne 0
\ee
This is, in fact, not a surprise, because the gradient of the rotated 
scale function $\tilde \lambda$ must be perpendicular to the rotated
vector $\vec {\tilde\Sigma}$, $(\nabla \tilde \lambda )\cdot \vec{\tilde
\Sigma}=0$, and analogously for other base space transformations.  

Therefore, in order to have a closing Lie algebra of base space and target
space transformations, we have to restrict to constant target space
scale transformations, $\lambda = {\rm const}$. 

However, as transformations with non-constant $\lambda$ obeying the constraint
(\ref{cond}) map solutions of the Dirac equation to new solutions and are of 
some 
interest as such, one may instead choose another restriction by restricting
to the target space transformations parametrized by $\lambda$ and $\phi$
(i.e., by setting all base space transformation parameters equal to zero).
The resulting Lie algebra is abelian and closes therefore trivially. 

We still have to calculate the coefficients $F^x ,F^y ,F^z$ from the
coefficients of the identity of Eqs. (\ref{pr-Deq-3d}). 
After a lenghty calculation
one obtains the simple result
\be \label{F-vec}
\vec F = - (\Theta +\sigma ) \vec G + \vec L \times \vec G -\nabla \phi
-\frac{1}{|\vec \Sigma |}\vec \Sigma \times \nabla \lambda
\ee

\section{Multiple zero modes}

In this section, we would like to apply the results on symmetries of Section
3 to the issue of zero
mode degeneracy (i.e., multiple solutions for a Dirac equation with
the same gauge potential). We shall give a more general discussion in
Subsection 4.1, whereas we will provide some explicit, new examples
of multiple zero modes in Subsection 4.2.

\subsection{General discussion}

The condition that a target space symmetry transformation of a spinor 
does not change the 
gauge field is $\vec F=0$, see Eq. (\ref{F-vec}), or 
\be \label{ph-la}
\nabla \phi =   -\frac{1}{|\vec \Sigma |}\vec \Sigma \times \nabla \lambda
\ee
which, together with condition (\ref{cond}), implies that $\vec \Sigma$,
$\nabla \phi$ and $\nabla \lambda$ are mutually perpendicular. Further,
Darboux's theorem tells us that (at least locally) we may express
$\vec \Sigma $ like
\be
\vec \Sigma = \nabla \xi^1 \times \nabla \xi^2
\ee
where $\xi^a =\xi^a ({\bf r})$, $a=1,2$. If we now assume that $\phi$ and
$\lambda $ are functions of $\xi^a$ only, then their gradients 
are automatically
perpendicular to $\vec \Sigma$, and it remains to solve
\be \label{ph-la-1}
| \nabla \phi | = |\nabla \lambda | \; , \quad \nabla \phi \cdot \nabla
\lambda =0.
\ee
This problem can be solved relatively easily for a subclass of functions
$\xi^a$ which fulfill one additional requirement, namely that the scalar
products of the gradients of the $\xi^a$ can be expressed in terms of
the $\xi^a$ again, up to a {\em common} factor, that is
\be \label{Ri-Su}
\nabla \xi^a \cdot \nabla \xi^b =h({\bf r}) g^{ab}(\xi^c).
\ee
Here $g^{ab}$ plays the role of a metric in some two-dimensional space
parametrized by the coordinates $\xi^a$.
If Eq. (\ref{Ri-Su}) holds, then Eq. (\ref{ph-la}) simplifies to
\be \label{ph-la-2}
\phi_a \tilde \epsilon^{ac} =g^{cb}\lambda_b
\ee
where $\phi_a \equiv \partial_{\xi^a} \phi$, etc. and
\be
\tilde \epsilon^{ab} =g^\frac{1}{2} \epsilon^{ab} \; ,\qquad
g\equiv {\rm det}(g_{ab}) =g_{11}g_{22}-g_{12}g_{21}
\ee
and $\epsilon^{ab}$ is the usual antisymmetric symbol in 2 dimensions.
Eqs. (\ref{ph-la-2}) can be solved by observing that they  just
provide a generalization of the Cauchy--Riemann equations to the case of a
general surface. We only have to find the coordinate transformation
from $\xi^a$ to some new coordinates ${\xi '}^a $ such that the metric
${g '}^{ab}$ w.r.t. the new coordinates is conformally equivalent to the
flat metric $\delta^{ab}$, that is
\be \label{g'}
{g '}^{ab} \equiv \frac{\partial {\xi '}^a}{\partial \xi^c}
 \frac{\partial {\xi '}^b}{\partial \xi^d} g^{cd} =h(\xi ') \delta^{cd}.
\ee
This problem always has a solution in two dimensions \cite{DFN}.
In terms of the coordinates ${\xi '}^a$ Eqs. (\ref{ph-la-2}) read 
\be \label{ph-la-3}
\phi_a  \epsilon^{ac} =\delta^{cb}\lambda_b
\ee
which are just the Cauchy--Riemann equations. Therefore, our problem is 
solved by an arbitrary complex function $u$ of the complex variable 
$\zeta ' $ where
\be \label{u}
u(\zeta ') =\exp (\lambda +i\phi ) \; , \quad \zeta '= {\xi '}^1 +i {\xi '}^2 .
\ee
Here ``solution'' means that if a spinor $\Psi$ with $\Psi^\dagger
\vec \sigma \Psi \equiv \vec \Sigma $ solves a certain Dirac equation,
then $u\Psi$ locally solves the same Dirac equation,
\be \label{u-psi}
{\cal D}\Psi =0 \quad \Rightarrow \quad {\cal D} u\Psi =0.
\ee
Regularity requirements further restrict the allowed $u$. For instance,
the functions $\xi^a$ need not be well-defined in all $\RR^3$ (they
usually are not), but $u$ certainly has to be well-defined
(which may not be possible, in which case the problem has no
acceptable solution). The condition
that $u\Psi$ is square-integrable 
restricts to a finite number (which, again, may be zero)
of linearly independent functions $u$. It should be emphasized that
{\em all} known examples of multiple zero modes  are of the above
type (\ref{u-psi}), see \cite{AMN3}--\cite{ES}. 
It is an interesting open question whether
there exist other types, as well.

Finally, let us point out that the procedure described above can certainly be
reversed. That is to say, choose a pair of functions $\xi^a$ which obey
Eq. (\ref{Ri-Su}) and calculate the corresponding ${\xi '}^a$ via
Eq. (\ref{g'}). Then $u\Psi$ with $u$ given by Eq. (\ref{u}) will
provide additional local zero modes for a whole class of spinors $\Psi$ with
spin densities given by $\vec \Sigma =F(\xi^a) \nabla \xi^1 \times 
\nabla \xi^2 $ for (almost) arbitrary real functions $f(\xi^a)$ (of course,
$f$ should be well-defined in all $\RR^3$). It is not difficult to 
reconstruct the spinor $\Psi$ from the spin density $\vec \Sigma$
and the gauge potential $\vec G$ from the spinor $\Psi$
(see, e.g., \cite{LY1}). We shall explicitly demonstrate how this works
by constructing a class of multiple zero modes in the next subsection.

[Remark: Equations (\ref{ph-la-1}) can be expressed in terms of the
complex variable $u=\exp (\lambda +i \phi )$ like
\be \label{eik}
(\nabla u)^2 =0
\ee
which is the complex static eikonal equation. Further, condition (\ref{Ri-Su})
is equivalent to the condition that $u$ - when interpreted as a map from
one-point compactified $\RR^3_0$ to one-point compactified $\CC_0$ -
is a Riemannian submersion up to Weyl transformations (local conformal
rescalings of the metric). That is to say, $u$ is a composition of maps
\be
u\; : \quad \RR^3_0 \; \stackrel{{\rm W}}{\ra} \; {\cal M}^3 \;
\stackrel{{\rm R. S. }}{\ra} \; {\cal N}^2 \stackrel{{\rm W}}{\ra}
\; \CC_0
\ee
where W is a Weyl transformation, R.S. is a Riemannian submersion, and
${\cal M}^3$ and ${\cal N}^2$ are compact manifolds in three and two
dimensions, respectively. A detailed discussion of these issues can be
found in \cite{eik}. In \cite{ES}, Riemannian submersions were used
to construct multiple zero modes within a 
more geometrical context.] 

\subsection{A class of multiple zero modes}

Here we want to construct explicitly a class of multiple zero modes by 
starting with a pair of functions $\xi^a$ which obey Eq. (\ref{Ri-Su}), as
explained in the last subsection. In particular, 
we want to make use of the results of \cite{eik}, where a class of Hopf maps 
obeying the eikonal equation (\ref{eik}) were found, and where the geometric 
explanation for their existence was provided. In concordance with these
results, we therefore choose the functions
\be \label{ho-var}
\xi^1 = \ln \sinh \eta \equiv \frac{1}{2} \ln T 
\, , \quad \xi^2 = m \vartheta + n \varphi
\ee
where we introduced toroidal coordinates $(\eta ,\vartheta ,\varphi )$
(and, for later convenience, the variable $T\equiv \sinh^2 \eta$)
related to the cartesian coordinates $(x,y,z)$ via
\be
T \equiv \sinh^2 \eta = \frac{4(x^2 +y^2)}{4z^2 + (r^2 -1)^2} \, , \quad
\vartheta = \arctan \frac{2z}{r^2 -1} \, ,\quad \varphi = \arctan
\frac{y}{x} .
\ee
Further, $m$ and $n$ are nonzero integers such that $\exp (2\pi il \xi^2 )$
is a single-valued function for integer $l$ (the geometric significance of
the integers $m$ and $n$ is that they provide the complex function
(Hopf map)
$f= \exp (\xi^1 + i \xi^2)$ with the Hopf index $H=mn$). 

Following the results of 
\cite{eik},  it can be shown that the pair $\xi^1 ,\xi^2$ obeys Eq.
(\ref{Ri-Su}), and that a new pair ${\xi '}^1 ,{\xi '}^2$ obeying
Eq. (\ref{g'}) can be found without difficulty. Indeed, one finds easily that
$\xi^a$ obey Eq. (\ref{Ri-Su}) with
\be
h({\bf r}) =\frac{(\cosh \eta -\cos \vartheta )^2 \cosh^2 \eta}{\sinh^2 \eta}
\ee
\be
g^{11}=1\, ,\quad g^{12} = g^{21}=0 \, ,\quad g^{22} = \frac{m^2 \sinh^2
\eta + n^2}{\cosh^2 \eta} .
\ee
Hence, the induced metric $g^{ab}$ is already diagonal but not yet 
conformally flat. However, as $g^{22}$ only depends on $\xi^1$ 
(i.e., on $\eta$), a coordinate transformation involving only $\xi^1$ is
sufficient, $\xi^1 \rightarrow {\xi '}^1 (\xi^1) $, ${\xi '}^2 \equiv \xi^2$,
such that $g^{11} \rightarrow {g'}^{11} =g^{22}\equiv {g'}^{22}$. 
Explicitly, the transformation reads
\be
{\xi'}^1 = \ln \left( 
 \sinh^{|n|} \eta \, \, 
\frac{\left( |m|\cosh \eta + \sqrt{n^2 +m^2 \sinh^2 \eta
}\right)^{|m|}}{\left( |n|\cosh \eta + \sqrt{n^2 +m^2 \sinh^2 \eta
}\right)^{|n|}} \right) ,
\ee
see \cite{eik}. The function $\zeta ' ={\xi '}^1 +i{\xi '}^2$ itself is not 
single-valued, but the function
\be
\zeta \equiv \exp \zeta ' = \exp ({\xi '}^1 +i{\xi '}^2 )
\ee
is. It follows that for each spinor $\Psi$ with spin density $ \vec \Sigma 
\equiv \Psi^\dagger \vec \sigma \Psi = F(\xi^a) \nabla \xi^1
\times \nabla \xi^2 $ (which is a formal zero mode for some gauge potential,
because $\vec \Sigma$ obeys $\nabla \cdot \vec \Sigma =0$), $u(\zeta)
\Psi$ are further formal zero modes for the same gauge potential, where
$u$ is a rational function of its argument. A customary basis for the
functions $u$ is $u=\zeta^l$ for integer $l$.  

The formal zero modes described so far (implicitly via their spin density) are 
single-valued functions on all $\RR^3$, but we have not yet taken into account
the condition of square-integrability.  Before doing so, we want to make 
some simplifying assumptions on the type of spin density we want to
discuss. Firstly, we assume that the function $F$ depends on $\xi^1$ 
(i.e., on $\eta$ or $T$) only. Secondly, we reexpress $F$ like
$F= e^M \frac{4T}{(1+T)^2}$, that is, we write for a general spin density
\be
\vec \Sigma^{(M)} = e^{M(T)}\vec \Sigma^{(0)} \, ,\quad
\vec \Sigma^{(0)}= \frac{4T}{(1+T)^2}\nabla \xi^1 \times \nabla \xi^2 .
\ee
The reason for this is that $\vec \Sigma^{(0)}$ is a well-behaved and 
integrable spin density, whereas $\nabla \xi^1 \times \nabla \xi^2$ is not
(clearly, integrability of the spin density $\int d^3 {\bf r} |\vec \Sigma |
< \infty $ is the same as square-integrability of the corresponding
spinor). Another reason for the choice of $\vec \Sigma^{(0)}$, which does
not concern us much here, is the fact that $\vec \Sigma^{(0)}$ is the
Hopf curvature for the Hopf map $\exp (\xi^1 +i\xi^2)$. In cartesian
coordinates, $\vec \Sigma^{(0)}$ reads
\be \label{Sig-0}
\vec \Sigma^{(0)} = \frac{16}{(1+r^2)^3} 
\left( \begin{array}{c} 2n x z -2m y  \\ 2n y z +2m x \\
n(1-r^2 +2 z^2) \end{array} \right) 
\ee
and its spinor $\Psi^{(0)}$ with $\vec \Sigma^{(0)}= \Psi^{(0) \dagger}
\vec \sigma \Psi^{(0)}$ is
\be \label{Psi-0}
\Psi^{(0)} = \frac{2\sqrt{2}}{(1+r^2)^{\frac{3}{2}}\sqrt{(1+r^2)\Gamma +
n(2z^2 +1-r^2)}} \cdot \nonu  
\ee
\be \label{spi-0}
 \cdot \left( \begin{array}{c} n(2z^2 +1-r^2) +(1+r^2)\Gamma  \\ 
2(x +iy) (nz +im) \end{array} \right) 
\ee
where
\be
\Gamma :=
\left( \frac{n^2 +m^2 T}{1+T} \right)^{\frac{1}{2}}
=\left( \frac{n^2 +m^2 \sinh^2 \eta}{\cosh^2 \eta} \right)^{\frac{1}{2}} 
\ee
and we chose the gauge such that the upper component is real. 
The calculation of the spinor $\Psi^{(0)}$ from the spin density
$\vec \Sigma^{(0)}$ is explained in Appendix B. The spinor
(\ref{spi-0}) is regular everywhere. Further, it is a zero mode 
for some gauge potential $\vec G^{(0)}$, by construction (this gauge 
potential is well-behaving and leads to a square-integrable magnetic
field $\vec H^{(0)} =\nabla \times G^{(0)}$; its explicit expression,
which is quite lengthy, is displayed in Appendix B, together with
an explanation of its calculation).
Additional formal zero modes $\zeta^l \Psi^{(0)}$
for the same gauge potential are not square-integrable, i.e., the
corresponding spin density $|\zeta^* \zeta|^l \vec \Sigma^{(0)}$ is not
integrable (for either positive or negative integer $l$), 
as may be inferred easily from the explicit expression for 
$\vec \Sigma^{(0)}$ and from the limiting behaviour
\be
\lim_{T\to 0} |\zeta^* \zeta| \sim T^{|n|} \, ,\quad 
\lim_{T\to \infty} |\zeta^* \zeta| \sim T^{|m|} .
\ee
But with an appropriate choice of $M$, it is easy to find spin densities
$ \vec \Sigma^{(M)}$ such that $|\zeta^* \zeta|^l \vec \Sigma^{(M)}$ remains
integrable for some non-zero values of $l$.

Concretely, we assume that $M(T)$ and $M'(T)$ are finite for 
all finite values of $T$, i.e.,
\be \label{M-fin}
|M(T)| < \infty \quad \wedge \quad |M' (T)| < \infty
\quad \mbox{for} \quad T < \infty ,
\ee
 and that $M$ behaves for large $T$ like  
\be \label{M-limit}
\lim_{T\to \infty} M(T) \sim -mk\ln T +\bar M(T)
\ee
where the remainder $\bar M$ has to obey
\be \label{M-rem}
\lim_{T\to \infty} |T\bar M(T)| < \infty .
\ee 
Here the conditions (\ref{M-fin}) and (\ref{M-rem}) are chosen in order 
to avoid both spurious (i.e., pure gauge) and physical singularities for
the corresponding gauge potential, and condition (\ref{M-limit}) is chosen
in order to have a non-trivial result (i.e., multiple square-integrable
zero modes). Condition (\ref{M-limit}) induces a spurious (gauge)
singularity in the corresponding gauge potential at $T=\infty$
which has to be cured by an appropriate gauge
fixing. Consequently, the spinor $\Psi^{(M)}$ for the spin density
$\vec \Sigma^{(M)}$ with the appropriate gauge fixing is
\be \label{M-mode}
\Psi^{(M)} :=e^{\frac{M}{2}}e^{i km \vartheta}\Psi^{(0)} 
\ee
where the additional factor $\exp (ikm \vartheta)$ provides the gauge fixing,
see below.
Under the above assumptions the spinors
\be \label{M-l-mode}
\Psi^{(M)}_l =\zeta^l \Psi^{(M)} \quad , \qquad l=0\ldots k
\ee
are all square-integrable 
zero modes for the same gauge potential $\vec G^{(M)}$.

It remains to determine $\vec G^{(M)}$ explicitly, which will be obtained
with Eq. (\ref{F-vec}) of Section 3. In fact, using Eq. 
(\ref{F-vec}) (where $ M/2 \sim \lambda$ and 
$km\vartheta \sim \phi$; remember that Eq. (\ref{F-vec}) also holds for finite
$\lambda$ and $\phi$ because of the abelian nature of the target space
symmetry transformations) one easily calculates
\be \label{G-M}
\vec G^{(M)} = \vec G^{(0)} 
-\frac{M' T}{\Gamma}(m\nabla \vartheta +n\nabla
\varphi ) - mk\nabla \vartheta 
\ee
(as said, the expression for $\vec G^{(0)}$ is provided in Appendix B).
Here the additional pure gauge term precisely cancels a pure gauge 
singularity at $T=\infty$, as may be checked easily. Further, we may see
how it works that multiplication of the spinor $\Psi^{(M)}$ by
$\zeta$ does not change the gauge potential. The crucial point is that
multiplication by $\zeta$ corresponds to choosing $M=|\zeta^* \zeta|$,
and for this $M$ it holds that $\frac{M' T}{\Gamma} =-1$, so that the
$M$ dependent term in  (\ref{G-M}) is, in fact, pure gauge and is
cancelled by the contribution $\nabla \arg \zeta =\nabla \xi^2$, see
(\ref{ho-var}). This just shows explicitly that $M/2 \equiv {\xi '}^1
\sim \lambda$ and ${\xi }^2 \sim \phi$ fulfill Eq. (\ref{ph-la}) which,
of course, must be true by construction.

Therefore, we have succeeded in constructing explicitly a class of
multiple zero modes together with their gauge potentials starting from
the functions (\ref{ho-var}) (or, equivalently, starting from the 
higher toroidal Hopf maps of Ref. \cite{eik}). For the simplest
Hopf map (i.e., for $m=n=1$) these zero modes have already been obtained in
\cite{AMN3,AMN4,ES}, whereas for the higher toroidal Hopf maps they are
new.  
\\ \\ \\ 
{\large\bf Appendix A} \\ \\
The coefficients for the prolongation ${\rm pr}\, v$ of the vector
field $v$ in Section 2 are
\br
R^x & =& D_x R -\rho_x D_x X - \rho_y D_x Y \\
R^y & =& D_y R -\rho_x D_y X - \rho_y D_y Y \\
L^x & =& D_x L -\lambda_x D_x X - \lambda_y D_x Y \\
L^y & =& D_y L -\lambda_x D_y X - \lambda_y D_y Y
\er
where $D_x$ and $D_y$ are total derivatives w.r.t. $x$ and $y$, 
respectively. Explicitly, $R^x$ reads
\br
R^x &=& R_x +R_\rho \rho_x +R_\lambda \lambda_x +R_{G^x} G^x_x +R_{G^y}
G^y_x \nonumber \\
& -&  \rho_x (X_x +X_\rho \rho_x +X_\lambda \lambda_x +X_{G^x}G^x_x
+X_{G^y} G^y_x ) \nonumber \\
& -& \rho_y (Y_x +Y_\rho \rho_x +Y_\lambda \lambda_x +Y_{G^x}G^x_x
+Y_{G^y} G^y_x )
\er
and analogously for the other coefficients.
\\ \\ \\ 
{\large\bf Appendix B} \\ \\
The calculation of the spinor $\Psi^{(0)}$ from the spin density $\vec 
\Sigma^{(0)}$ proceeds as follows. For a gauge such that the upper component
of the spinor is real, a spinor $\Psi$ may be obtained from its spin density
$\vec \Sigma$ formally by the algebraic relation, see Ref. \cite{LY1},
\be
\Psi = \frac{1}{\sqrt{2(|\vec \Sigma | +\Sigma^3 )}} 
\left( \begin{array}{c} |\vec \Sigma | +\Sigma^3  \\ 
\Sigma^1 + i\Sigma^2 \end{array} \right) .
\ee
This gauge is globally
admissible provided that the third component of $\vec \Sigma$
is positive whenever its first and second component are zero, i.e.,
\be
\Sigma^1 =0 \, \wedge \, \Sigma^2 =0 \quad \Rightarrow \quad \Sigma^3 >0
\ee
which holds for the spin density
$\vec \Sigma^{(0)}$ of Eq. (\ref{Sig-0}). Therefore,
the spinor $\Psi^{(0)}$ of Eq. (\ref{Psi-0}) may be computed with the help 
of the above expression.

The calculation of the gauge potential $\vec G^{(0)}$ for the spinor
(zero mode) $\Psi^{(0)}$ proceeds as follows. Remember that once a zero mode
$\Psi$ is given, the Dirac equation may be viewed as an overconstrained system
of linear, algebraic equations for the three components of the gauge
potential. If the constraint is fulfilled, this system can be solved
explicitly and results in the following expression for the corresponding
gauge potential $\vec G$, see Ref. \cite{LY1},
\be \label{form-G}
\vec G = \frac{1}{2 |\vec \Sigma |} \left( \nabla \times \vec \Sigma +
2 \, {\rm Im} \, ( \Psi^\dagger \nabla \Psi ) \right) .
\ee
Inserting the explicit expression for $\Psi^{(0)}$ into the above formula
leads to the following rather lengthy expression for $\vec G^{(0)}$,
\be
\vec G^{(0)} = \frac{1}{\Gamma (1+r^2 )^2} \left( \begin{array}{c}
-n(5-r^2) y +6mxz \\ n(5-r^2) x + 6myz \\ m(6z^2 +2 -4r^2) \end{array}
\right) + \nonumber
\ee
\be
+ \frac{2}{\Gamma (1+r^2)[ \Gamma (1+r^2) +n(2z^2 +1 -r^2)]} \left(
\begin{array}{c} -(n^2 z^2 +m^2)y \\ (n^2 z^2 +m^2 )x \\ mn (r^2 -z^2)
\end{array}  \right)
\ee
(for a direct comparison with the results of \cite{AMN3,AMN4} for the case 
$m=n=1$, one must take into account the fact that the gauge chosen in 
\cite{AMN3,AMN4} differs from the one chosen here by the gauge function
$\phi =\arctan (z)$).
Further, Eq. (\ref{form-G}) shows that $\vec G$ is singular whenever
$|\vec \Sigma|$ is zero. This singularity may be a spurious (gauge) 
singularity, in which case it can be cured by an
appropriate gauge fixing as in Eq. (\ref{G-M}), or it may be physical, in
which case the corresponding zero mode is not admissible (of course, all the
zero modes given in Subsection 4.2 are admissible).
\\ \\ \\
{\large\bf Acknowledgement:} \\
This research was partly supported by MCyT(Spain) and FEDER
(FPA2002-01161), Incentivos from Xunta de Galicia and the EC network
"EUCLID". Further, CA acknowledges support from the 
Austrian START award project FWF-Y-137-TEC  
and from the  FWF project P161 05 NO 5 of N.J. Mauser.

\end{document}